\documentclass[prl,aps,10pt,twocolumn,superscriptaddress]{revtex4-1}
\usepackage{epsfig,amsmath,amssymb,graphicx,color,calc}

\newcommand{\beq}{\begin{equation}}
\newcommand{\eeq}{\end{equation}}
\newcommand{\ba}{\begin{align}}
\newcommand{\ea}{\end{align}}
\renewcommand{\phi}{\varphi}

\begin{document}

%%%%%%%%%%%%%%%%%%%%%%%%%%%%%%

%% For titles, only capitalize the first letter
%% \title{Almost sharp fronts for the surface quasi-geostrophic equation}

\title{Dimensional Study of the Caging Order Parameter at the Glass Transition}

%% Enter authors via the \author command.  
%% Use \affil to define affiliations.
%% (Leave no spaces between author name and \affil command)

%% Note that the \thanks{} command has been disabled in favor of
%% a generic, reserved space for PNAS publication footnotes.

%% \author{<author name>
%% \affil{<number>}{<Institution>}} One number for each institution.
%% The same number should be used for authors that
%% are affiliated with the same institution, after the first time
%% only the number is needed, ie, \affil{number}{text}, \affil{number}{}
%% Then, before last author ...
%% \and
%% \author{<author name>
%% \affil{<number>}{}}

%% For example, assuming Garcia and Sonnery are both affiliated with
%% Universidad de Murcia:
%% \author{Roberta Graff\affil{1}{University of Cambridge, Cambridge,
%% United Kingdom},
%% Javier de Ruiz Garcia\affil{2}{Universidad de Murcia, Bioquimica y Biologia
%% Molecular, Murcia, Spain}, \and Franklin Sonnery\affil{2}{}}

\author{Patrick Charbonneau}
\affiliation{Departments of Chemistry and Physics, Duke University, Durham,
North Carolina 27708, USA}

\author{Atsushi Ikeda}
\affiliation{Laboratoire Charles Coulomb,
UMR 5221 CNRS and Universit\'e Montpellier 2, Montpellier, France}

\author{Giorgio Parisi}
\affiliation{Dipartimento di Fisica,
Sapienza Universit\'a di Roma,
INFN, Sezione di Roma I, IPFC -- CNR,
P.le A. Moro 2, I-00185 Roma, Italy
}

\author{Francesco Zamponi}
\affiliation{LPT,
\'Ecole Normale Sup\'erieure, UMR 8549 CNRS, 24 Rue Lhomond, 75005 France}

%\contributor{Submitted to Proceedings of the National Academy of Sciences
%of the United States of America}

%% The \maketitle command is necessary to build the title page.

%%%%%%%%%%%%%%%%%%%%%%%%%%%%%%%%%%%%%%%%%%%%%%%%%%%%%%%%%%%%%%%%
%\begin{article}

\begin{abstract} The glass problem is notoriously hard and controversial.
Even at the mean-field level, little is agreed about how a fluid turns sluggish while exhibiting but unremarkable structural changes. It is clear, however, that the process involves self-caging, which provides an order parameter for the transition. %~\cite{berthier:2011}. 
It is also broadly assumed that this cage should have a Gaussian shape in the mean-field limit. Here we show that this ansatz does not hold. By performing simulations as a function of spatial dimension $d$, we find the cage to keep a non-trivial form. Quantitative mean-field descriptions of the glass transition, such as mode-coupling theory, %~\cite{gotze:2009}, 
density functional theory, %~\cite{kirkpatrick:1987}, 
and replica theory, %~\cite{MP09,PZ10}, 
all miss this crucial element. Although the mean-field random first-order transition %~\cite{kirkpatrick:1989} 
scenario of the glass transition is here qualitatively supported and non-mean-field corrections are found to remain small on decreasing $d$, 
reconsideration of its implementation is needed for it to result in a coherent description of experimental observations.
\end{abstract}

\maketitle
%% When adding keywords, separate each term with a straight line: |
%\keywords{glass transition| mean-field theory | order parameter}

%% Optional for entering abbreviations, separate the abbreviation from
%% its definition with a comma, separate each pair with a semicolon:
%% for example:
%% \abbreviations{SAM, self-assembled monolayer; OTS,
%% octadecyltrichlorosilane}

% \abbreviations{}

%% The first letter of the article should be drop cap: \dropcap{}
%\dropcap{I}n this article we study the evolution of ''almost-sharp'' fronts

%% Enter the text of your article beginning here and ending before
%% \begin{acknowledgements}
%% Section head commands for your reference:
%% \section{}
%% \subsection{}
%% \subsubsection{}

\paragraph*{Introduction -}
If crystallization is avoided, slowly compressed (or supercooled) fluids eventually form a glass. They become non-ergodic when their structural relaxation timescale $\tau_{\alpha}$ (or inverse diffusivity $1/D$) gets larger than the annealing time. 
A variety of competing descriptions propose to explain this seemingly straightforward process~\cite{BB11}, but existing experimental and numerical results do not allow to unambiguously discriminate between them. Yet consensus has recently emerged that a growing dynamical length scale is associated with the transition, which some have argued results 
in a critical phenomenon of a new kind~\cite{berthier:2011}.
Based on this development, 
it seems natural to rephrase the problem starting from a mean-field (MF) theory, in which correlations are neglected, 
and to add correlations progressively using renormalization group techniques.
Unfortunately, even identifying what should be the MF microscopic phenomenology is fraught with contention. 
One of the common MF framework is the random first-order transition (RFOT) theory, which stems from the exact solution of a class of abstract spin glass models whose phenomenology is remarkably similar to that of structural glass formers~\cite{kirkpatrick:1988,kirkpatrick:1989}. Inspired by this analogy, reasonable predictions have been obtained for realistic models~\cite{LW07,MP09,PZ10}. 
Yet despite these advances, the foundations of the RFOT scenario are insufficiently robust for it to be widely accepted as {\it the} MF theory of glasses, leaving ample room for criticism and alternative formulations~\cite{BB11,berthier:2011,Ca09,BB09,keys:2011}.

Briefly, the RFOT scenario states that an unavoidable ergodicity breaking occurs at  finite pressures and temperatures, whatever the annealing rate. 
In the MF approximation, which is assumed to hold when $d\rightarrow\infty$,
$\tau_{\alpha}$
diverges at a dynamical transition associated with self-caging at which phase space breaks up into pure states. 
In finite dimensions, the growth of $\tau_{\alpha}$ at the dynamical transition is limited, because nucleation of one glassy state from another may be possible up to the Kauzmann transition, 
beyond which only the lowest free-energy state prevails~\cite{kirkpatrick:1989}. 
But because nucleation gets strongly
suppressed with $d$, the dynamical transition is thought to dominate the slowdown in high $d$. 

Like for critical phenomena, RFOT's MF predictions are expected to be  more
accurate above an upper critical
dimension, shown to be $d_u = 8$~\cite{biroli:2007,FPRR11}. On finite-dimensional lattices, a renormalization group analysis has even found a RFOT-like fixed point~\cite{CBTT11}, although in certain cases it disappears in low $d$~\cite{YM12}.
Given that these analyses are restricted to abstract models,
a key question is whether RFOT provides reliable {\it quantitative}
predictions for a realistic particle model, at least in large $d$.
This program was initiated soon after the theory's formulation~\cite{kirkpatrick:1987}, but is no simple task. 
The key difficulty is that, while for spin glass models the order parameter for the glass transition is the relatively straightforward local Edwards-Anderson overlap~\cite{kirkpatrick:1989},
for particle systems the caging order parameter is a non-trivial function of space, the so-called {\it non-ergodic parameter} or its Fourier transform, the van Hove function $G_\mathrm{s}(r,t)$~\cite{gotze:2009}.
The RFOT formulation for particle-based systems requires a set of integral equations to describe $G_\mathrm{s}(r,t)$, which is only achieved under 
 poorly controlled approximations and results in non-equivalent treatments in finite $d$.
Most of these formulations, such as density-functional theory (DFT)~\cite{kirkpatrick:1987} and replica theory (RT)~\cite{MP09,PZ10}, can be extended to $d\rightarrow\infty$, so if their underlying approximations were truly 
MF in nature they should provide equivalent and accurate predictions in that limit. 
Although mode-coupling theory (MCT) was developed independently from RFOT~\cite{gotze:2009}, 
many have suggested that a dynamical description of the RFOT scenario should result in MCT-like equations~\cite{kirkpatrick:1987,andreanov:2009}.
This observation raised the question whether MCT
should converge to the correct mean-field description in high $d$~\cite{kirkpatrick:1987}.
MCT's results for the glass transition were, however, recently found to be not only asymptotically divergent from DFT's and RT's~\cite{ikeda:2010,schmid:2010}, 
but to get increasingly unphysical even in relatively low dimensions~\cite{charbonneau:2011}. In order to evaluate the  MF scenario for the glass transition, we 
here stringently test these theories against simulation results as a function of $d$, emphasizing the evolution of the caging order parameter.

\begin{figure}
\includegraphics[width=\columnwidth]{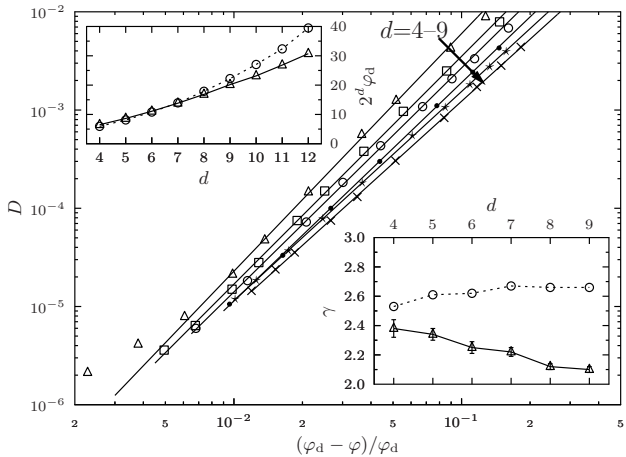}
\caption{
Power-law fits (lines) to the vanishing diffusivity improve with $d$=4--9 (different symbols), spanning over three decades of $D$. (top inset) The resulting numerical $\phi_{\mathrm{d}}$ values (including the $d=10-12$ results from Ref.~\cite{charbonneau:2011}) are, however, significantly different from the MCT results for the dynamical transition (short-dash line). (bottom inset) The results for $\gamma$ also disagree with MCT predictions.
}
\label{fig:gamma}
\end{figure}

\begin{figure}
\includegraphics[width=\columnwidth]{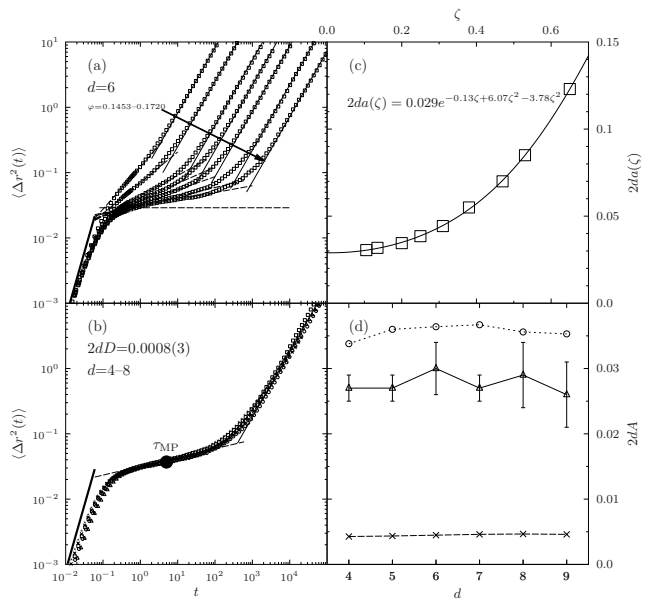}
\caption{
(a) The MSD in $d=6$ for increasing $\phi$=0.1453--0.1720 illustrates the developing caging regime (dashed line), intermediate between the ballistic (thick line) and the diffusive (solid line) regimes. (b) The MSD for isodiffusive states in $d$=4--8 identifies the caging mid-point time $\tau_{\mathrm{MP}}$ (large dot). The power law fitting parameters for the caging regime in $d=6$ from (a) are used in (c) to extract the plateau height at the dynamical transition, when $\zeta=0$. (d) The plateau height (solid line) is consistent with RT's lower bound at $\phi_{\rm K}$ (long-dash line) and significantly different from the MCT predictions (short-dash line).
}
\label{fig:MSD}
\end{figure}

\begin{figure}
\includegraphics[width=\columnwidth]{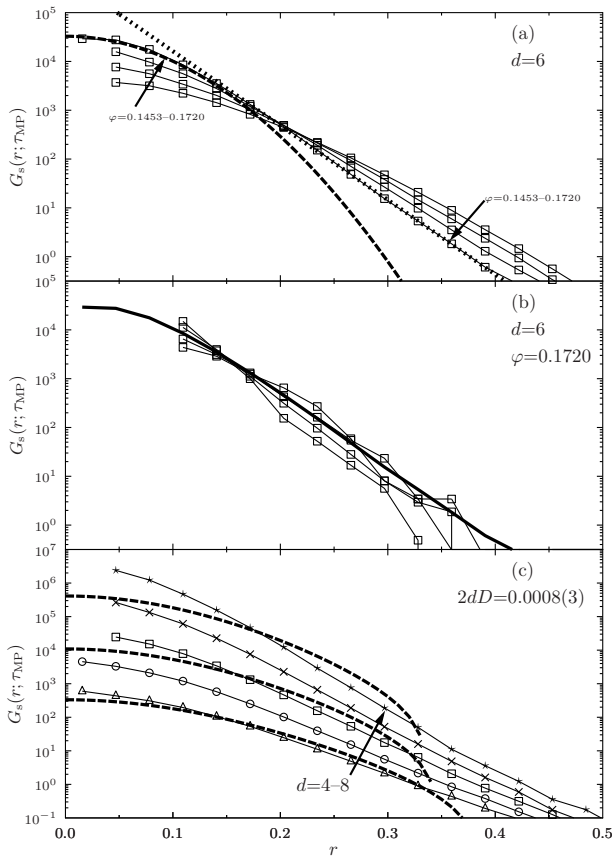}
\caption{
(a) The evolution of the van Hove function with packing fraction at $\tau_{\mathrm{MP}}$ in $d=6$ shows that the fat exponential tail (short-dash line) steadily grows at the expense of the Gaussian regime (long-dash line). (b) The isoconfigurational results for four randomly chosen particles (symbols) at $\phi=0.1720$ in $d=6$ indicate that the individual cages as well as the average cage (thick line) are non-Gaussian. (c) The isodiffusivity indicates that the fat tail remains undiminished for all the dimensions studied, and grows increasingly different from the MCT results (long-dash lines), given here for $d$=4, 6, and 8.
}
\label{fig:cage}
\end{figure}

\begin{figure}
\includegraphics[width=\columnwidth]{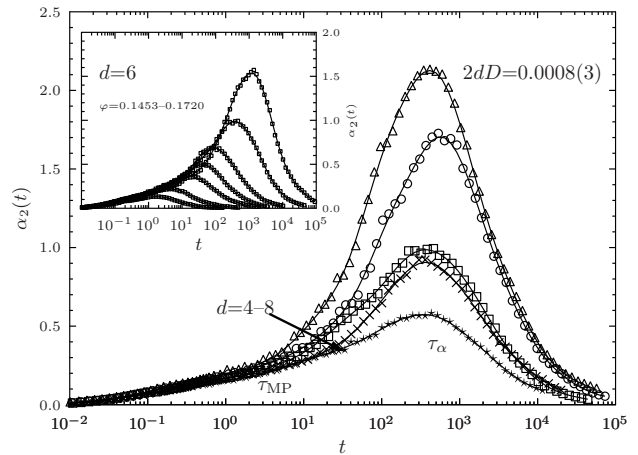}
\caption{The isodiffusive for $d=$4--8 (main)  and $d=6$ with increasing $\phi$ (inset) non-Gaussian parameter $\alpha_2(t)$. The peak at $\tau_\alpha$ decays strongly with $d$, an indication that dynamical heterogeneity is increasingly suppressed, while the caging behavior near $\tau_{\mathrm{MP}}$ is robust.
}
\label{fig:alpha2}
\end{figure}

\paragraph*{Results - }
Hard sphere fluids are the simplest glass formers with which to compare theoretical prediction, because their structure gets increasingly trivial with $d$~\cite{FP99,PS00}. As a first control, we test MCT's power-law scaling form for the vanishing diffusivity $D\sim(\phi-\phi_\mathrm{d})^{\gamma}$ when the fluid packing fraction $\phi$ approaches the dynamical transition at $\phi_{\mathrm{d}}$. 
This form 
fits high-$d$ results well, except for the more sluggish systems in $d=4$  (Fig.~\ref{fig:gamma})~\cite{charbonneau:2010},
and the $\phi_{\mathrm{d}}$ values agree with those obtained from a different procedure~\cite{charbonneau:2011}. 
MCT predicts values for $\phi_{\mathrm{d}}$ and $\gamma$, however, that are inconsistent with the numerical results, while
RT predicts a Kauzmann transition bound $\phi_{\rm K}>\phi_{\mathrm{d}}$ consistent with the numerical data~\cite{charbonneau:2011}, but results for $\gamma$ are still missing~\cite{franz:2012}.

The caging order parameter at the dynamical transition should offer a clearer picture of what is happening. In the high-$d$ MF limit, the van Hove function is argued to be Gaussian based on the multiplicity of ``uncorrelated'' caging neighbors and the central-limit theorem. 
In practice, most implementations of both RT~\cite{MP09,PZ10} and DFT~\cite{kirkpatrick:1987} simply assume a Gaussian form; while MCT predictions have been understood as faulty partly because they do not tend toward one~\cite{ikeda:2010}. 
Belief in the Gaussian form is so anchored that sustained deviations from it were quickly interpreted as dynamical heterogeneity absent from the 
MF picture~\cite{berthier:2011,kob:1997,HH96}. Directly measuring the van Hove function for hard spheres at $\phi_{\mathrm{d}}$, i.e., once diffusion is fully suppressed, is, however, impossible. In low $d$, the dynamical transition from the RFOT scenario is avoided, which transforms the arrest into a dynamical crossover and blurs its properties. Although increasing dimensionality resolves this ambiguity, reaching equilibrated configurations near $\phi_{\mathrm{d}}$ is challenging. Annealing more slowly than $\tau_{\alpha}$ limits the numerically accessible $\tau_{\alpha}$ to those within a few orders of magnitude from the collision time.
%Unlike for crystals or quasicrystals, where a nearly equilibrated configuration can be obtained from simple generating rules, or for certain models with quenched disorder, where planting configurations circumvents the issue~\cite{krzakala:2009}, no expedient point process generates dense fluid configurations. 

To circumvent this difficulty, we examine the systematic development of the caging regime of the mean square displacement (MSD) when approaching the dynamical arrest as $d$ increases. 
Near $\phi_{\mathrm{d}}$ %, in qualitative agreement with MCT predictions,
the MSD develops an inflection between the ballistic and the diffusive regimes,  which should plateau at full caging (Fig.~\ref{fig:MSD}a). 
On a purely phenomenological basis, we describe this intermediate regime by a power law
\begin{equation}
\langle \Delta r^2(t)\rangle = 2 d \, a(\phi) \, t^{\zeta(\phi)} \ ,
\end{equation}
whose subdiffusive exponent $\zeta(\phi)<1$ decreases with increasing $\phi$. Under the reasonable assumption that $\zeta(\phi) \to 0$ for $\phi\to\phi_{\rm d}$,
extrapolating the parametric plot $a(\zeta)$ to the limit $\zeta\to 0$ 
gives the cage size $A = a(\zeta\to 0)$ (Fig.~\ref{fig:MSD}c).
This measure is found to remain essentially constant $2dA \approx 0.027(4)$ over the $d$ range considered.  The MCT description of the intermediate regime is in qualitative, although not quantitative, agreement, and RT's lower bound from caging at $\phi_{\mathrm{K}}$ is respected (Fig.~\ref{fig:MSD}d).

Considering the caging order parameter, rather than the cage size,  more directly probes glass formation. We consider the evolution of $G_\mathrm{s}(r,\tau_{\mathrm{MP}})$ at the logarithmic mid-point $\tau_{\rm MP}$ 
of the intermediate caging regime. This choice has the advantage of correctly extrapolating to the full caging limit at the dynamical arrest, while staying well clear of $\tau_{\alpha}$ (Fig.~\ref{fig:MSD}b). The results surprisingly indicate that the cage shape does not tend toward a Gaussian (Fig.~\ref{fig:cage}). 
The Gaussian regime in fact shrinks to smaller $r$ with increasing $d$ and $\phi$, leaving instead a remarkably fat tail. %The isodiffusive development of the cage with dimension indicates that this tail remains prominent even in $d=8$. 
This result  contrasts with the RT and DFT assumptions, and markedly differs from the MCT predictions, whose discrepancy grows worse with $d$.

One may wonder if this pronounced deviation from the expected Gaussian behavior is due to dynamical heterogeneity, and to the growth of an associated dynamical length scale, 
as has been found in low $d$~\cite{berthier:2011,keys:2011,widmer-cooper:2004,chaudhuri:2007,lechenault:2010}.
The RFOT scenario and a description based on dynamical facilitation~\cite{keys:2011,garrahan:2003,hedges:2009} both suggest that such a length scale should be present in all $d$, but like other critical lengths, the distance $(\phi-\phi_{\rm d})/\phi_{\rm d}$ 
from the critical point over which it is felt is expected to shrink with $d$~\cite{biroli:2007,FPRR11}.  
The impact of dynamical heterogeneity should thus effectively disappear with increasing $d$.
The van Hove function unambiguously resolves the two processes.
We first perform an isoconfigurational study, in which a same initial configuration is randomly assigned series of different random initial momenta~\cite{widmer-cooper:2004}, indicating that the individual particle cages are not Gaussian either (Fig.~\ref{fig:cage}b). If the non-Gaussianity arose from an heterogeneity of the local relaxation on the $\tau_{\mathrm{MP}}$ scale, then one would expect the individual cages to be Gaussian, which they are not.
We then consider the non-Gaussian parameter $\alpha_2(t)$, which is the kurtosis of $G_\mathrm{s}(r,t)$ (Fig.~\ref{fig:alpha2}). Although $\alpha_2(t)$ decreases with $d$ for all time regimes, for isodiffusive systems 
the change is much more pronounced at the peak of $\alpha_2(t)$ 
near $\tau_\alpha$, where dynamical heterogeneity is maximal, than at the caging mid-point $\tau_{\mathrm{MP}}$. 
These results therefore support the numerical evidence that 
$G_\mathrm{s}(r,\tau_{\mathrm{MP}})$ remains non-Gaussian in the MF large $d$ caging regime.

\paragraph*{Discussion - }
Our work clarifies the MF scenario of the glass transition and establishes mileposts for assessing current and future theoretical descriptions of the phenomenon. The results suggest that the RFOT scenario qualitatively describes high-dimensional hard spheres, 
and that non-MF corrections remain small upon
decreasing dimension, even below $d_u$. 
When $d$ increases, the power-law divergence of $1/D$
near $\phi_{\rm d}$ is clearly visible; the associated dynamical heterogeneities  around $\tau_\alpha$ meanwhile decrease, making the glass transition  a {\it local} caging problem describable by MF theory. Yet, contrary to common belief, local caging does not lead to a simple Gaussian caging order parameter. 

We find a smooth dimensional dependence of the structural and dynamical properties, which is consistent
with what is found in the dynamical facilitation scenario~\cite{ashton:2005}. 
One might thus wonder whether a smooth $d$ dependence disagrees with the RFOT picture, which is based 
on an underlying critical phenomenon with an associated upper critical dimension $d_u = 8$~\cite{biroli:2007,FPRR11}.
Yet as in standard critical phenomena, the Landau-Ginzburg criterion indicates that one should be extremely close to
the critical point to see deviations from the MF predictions, even below $d_u$.
Recent quantitative computations show that the regime where non-MF corrections are present is quite hard to access using
numerical simulations in $d=3$~\cite{BB09,franz:2012}, and it is reasonable to argue that it should be even harder to reach with increasing $d$.
The existence of $d_u$ is therefore expected to be undetectable unless one is much closer to the critical point than we are here, and our results thus remain qualitatively consistent with the RFOT scenario.

Despite this qualitative agreement, we show that all the concrete implementations of RFOT theory struggle to describe the high $d$ regime, although it should be the easiest.
We find that a broad scope of MCT predictions are defective: the predictions for $\phi_{\rm d}$, the exponent $\gamma$, and the cage shape are not only wrong, but worsen with increasing $d$. These results reveal the inadequacy of standard MCT as a MF description, challenge some of the deep-seated 
assumptions about glass formation, and strongly call for a revised formulation of a dynamical theory of the RFOT~\cite{charbonneau:2011}.
At the same time, DFT and RT assume from the very beginning a Gaussian form for the cage, which is invalidated by our results. Even if some of the RT results seem consistent with our computations,
the theory should also be
revised to understand the extent to which a non-Gaussian caging order parameter affects its predictions.

If the RFOT formulation is indeed correct, we expect that theoretical reconsiderations will lead to a resolution of the discrepancy between
the MCT and RT/DFT predictions~\cite{ikeda:2010,schmid:2010},
and ultimately to a consistent description of
both the statics and dynamics of glass formers. 
Given that non-MF corrections seem small even for $d<d_u$, it is even possible
that such a complete MF theory could perform quite well in experimentally relevant dimensions. It would therefore  provide a productive starting point for a more refined renormalization
group analysis~\cite{CBTT11} that takes into account the role of fluctuations below $d_u$. It would also be interesting to evaluate the high-dimensional robustness of descriptions based on dynamical facilitation~\cite{garrahan:2003,hedges:2009,keys:2011}.
This program seems a promising route for obtaining a more robust and less controversial theory of glasses.

%% == end of paper:

%% Optional Materials and Methods Section
%% The Materials and Methods section header will be added automatically.

%% Enter any subheads and the Materials and Methods text below.
%\begin{materials}

\begin{table}
\caption{Numerical properties extracted from simulations.}
\label{table:fit}
%\begin{tabular*}{\hsize}{@{\extracolsep{\fill}}c c c c }
\begin{tabular}{@{\vrule height 10.5pt depth4pt  width0pt}cccc}
%\begin{tabular}{c c c c }
\hline 
$d$ & $\phi_\mathrm{d}$ & $\gamma$ & $A$\\ 
%\hline
%3 & 0.5950 & 2.42(5) & n/a\\
\hline 
4 & 0.4065 & 2.38(6) & 0.027(2) \\ 
\hline 
5 & 0.2700 & 2.34(4) & 0.027(2)\\ 
\hline 
6 & 0.1732 & 2.25(4) & 0.030(4) \\ 
\hline 
7 & 0.1081 & 2.22(3) & 0.027(2)\\ 
\hline 
8 & 0.06583 & 2.12(3) & 0.029(5)\\ 
\hline 
9 &  0.03938 &2.10(3) & 0.026(5)\\ 
\hline 
\end{tabular} 
\end{table}

\section{Materials and Methods}
\subsection{Numerical Simulations}
Event-driven molecular dynamics simulations of 8000 hard spheres in dimensions $4\leq d\leq9$ are performed under periodic boundary conditions~\cite{skoge:2006,charbonneau:2011}.  Because crystallization in high $d$ is strongly suppressed, access to deeply supersaturated starting configurations can be achieved via the slow compression of a low-density fluid~\cite{skoge:2006,vanmeel:2009b}. Between 4 and 8 independent configurations are equilibrated for each packing fraction. Simulations are then run at constant unit temperature $k_BT$ for times at least as large as $10/(2dD)$, where time $t$ is expressed in units of $( m\sigma^2/k_BT)$ for particles of unit mass $m$ and unit diameter $\sigma$. Even in $d=9$ for the densest system studied, the box side is kept greater than $2\sigma$. Because the static and dynamical correlation lengths shrink with $d$, these system sizes avoid significant finite-size effects,  as discussed in~\cite{charbonneau:2011}.

The average mean-square displacement (MSD) 
\begin{equation}\langle \Delta r^2(t)\rangle = N^{-1} \sum_{i=1}^N \langle |\textbf{r}_i(t)-\textbf{r}_i(0)|^2 \rangle
\end{equation}
is obtained from equilibrated starting configurations. At times shorter than the collision time, MSD displays a ballistic regime $\langle \Delta r^2(t)\rangle=dt^2$, and at long times it has a diffusive regime $\langle \Delta r^2(t)\rangle=2dDt$.

Fitting these numerically determined $D$ to the %RFOT 
power-law form $D\sim(\phi-\phi_\mathrm{d})^{\gamma}$
is reasonably good for $D<0.005$, and improves with increasing $d$. The resulting values of $\gamma$ and $\phi_{\rm d}$ are reported in table~1
%\ref{table:fit} 
and Fig.~\ref{fig:gamma}.
For $d\geq5$~\cite{charbonneau:2010}, the full accessible dynamical range studied is used, spanning up to three $D$ decades. Isodiffusive comparisons are made for systems whose $\phi$ gives $2dD= 0.0008(3)$. Because $\gamma$ differs relatively little over the $d$ range studied, choosing isodiffusive systems is roughly equivalent to keeping the distance to the dynamical transition $(\phi-\phi_\mathrm{d})/\phi_\mathrm{d}$ constant.
Note that small differences in $2dD$ can affect some measures, such as the non-smooth evolution with $d$ of the peak near $\tau_\alpha$ 
in Fig.~\ref{fig:alpha2}.

The cage is described by the self part of the van Hove function 
\begin{equation}
G_\mathrm{s}(r;t)=\frac{1}{N}\sum_{i=1}^N\langle\delta(|\textbf{r}_i(t)-\textbf{r}_i(0)|-r)\rangle \ ,
\end{equation}
which in the ballistic and diffusive regimes is well-approximated by a pure Gaussian.  The logarithmic caging mid-point  $\tau_{\mathrm{MP}}$ is chosen at the mid-time on a logarithmic scale, intermediate between the ballistic and the diffusive extrapolations of the MSD. The isoconfigurational study of the cage was repeated for 1000 different random initial velocity distributions, in order to obtain a good statistics on the individual cages.
The non-Gaussian character of this distribution is canonically described by its kurtosis, or non-Gaussian parameter,
\begin{equation}
\alpha_2(t)=\frac{d}{d+2}\frac{\langle \Delta r^4(t)\rangle}{\langle \Delta r^2(t)\rangle^2}-1 \ .
\end{equation}

\subsection{Mode-Coupling Theory}
The MCT analysis follows the approach of Refs.~\cite{ikeda:2010,schmid:2010}, using the Percus-Yevick (PY) structure factor calculated iteratively with a numerical Hankel transformation of order $d/2-1$. The agreement between the structural PY prediction and the numerical results improves with $d$, so the MCT predictions are not expected to depend on this choice of input. Using the hypernetted chain (HNC) input for the structure factor only worsens the agreement with simulations.
The long time limit of the self part of the van Hove function $G_\mathrm{s}(r;t\rightarrow\infty)$ is calculated 
by Fourier transforming the MCT solution of the self part of the non-ergodic parameter $f_\mathrm{s}(\vec{q})$. 
The plateau height at the long time limit of the MSD at caging $2dA= \lim_{t \rightarrow\infty} \langle\Delta r^2(t)\rangle$ 
is calculated through a small wave number analysis of the MCT equation,  
\begin{eqnarray}
A^{-1} = \frac{\rho}{d} \int \frac{d\vec{q}}{(2\pi)^d} q^2 c(\vec{q})^2 S(\vec{q}) f(\vec{q}) f_\mathrm{s}(\vec{q}), 
\end{eqnarray}
where $f(\vec{q})$ is the collective part of the non-ergodic parameter, for a given static structure factor $S(q)$ and direct correlation function $c(q)$.
In order to check the consistency of the numerical calculation, 
we also obtained the plateau height through $2dA= \int d\vec{r} \ r^2 G_\mathrm{s}(r;t\rightarrow\infty)$ and 
found the relative error to be smaller than 1\%.

\subsection{Replica Theory}

The best replica scheme for studying hard spheres is the small cage expansion~\cite{PZ10}. 
It consists of taking the lowest-order expansion of the replica theory free energy in the cage size $A$, 
as given in Ref.~\cite[Section VII, Eq.(73)]{PZ10}.
This approximation gives reliable results at high density near jamming, where $A$ is small.
Unfortunately, using this scheme the dynamical transition, which corresponds to the point where the self-consistent 
solution for states with cage $A$ vanishes, is not found. Because of the crudeness of this approximation, 
the equation for $A$ has indeed a solution $A^*(\phi)$ at all
$\phi$~\cite[Eq.(74)]{PZ10}.
Taking into account higher orders in the small $A$ expansion is only possible in the limit $d\to\infty$,
leading to an asymptotic prediction for the dynamical transition~\cite{PZ10}. However, this asymptotic limit
is reached only for extremely high dimensions, $d \gtrsim 50$, that are not accessible to simulations~\cite{charbonneau:2011}.
Other replica schemes are available~\cite{PZ10}, but they do not give good quantitative results. 
In summary,
for the moment replica theory does not give reliable predictions 
in the regime that is relevant for the present study,
namely $\phi\sim \phi_{\rm d}$ and low $d$. 
Using the lowest order expansion in $A$~\cite[Section VII, Eqs.(73)]{PZ10},
we can nonetheless obtain the cage radius at the Kauzmann transition $\phi_{\rm K} > \phi_{\rm d}$, which provides a lower bound $A^*(\phi_{\rm d}) > A^*(\phi_{\rm K})$ for the cage size.

%% Optional Appendix or Appendices
%% \appendix Appendix text...
%% or, for appendix with title, use square brackets:
%% \appendix[Appendix Title]

\begin{acknowledgments}
We thank J. Kurchan and R. Schilling for stimulating discussions. P.C. acknowledges NSF support No. DMR-1055586.
\end{acknowledgments}

%% PNAS does not support submission of supporting .tex files such as BibTeX.
%% Instead all references must be included in the article .tex document. 
%% If you currently use BibTeX, your bibliography is formed because the 
%% command \verb+\bibliography{}+ brings the <filename>.bbl file into your
%% .tex document. To conform to PNAS requirements, copy the reference listings
%% from your .bbl file and add them to the article .tex file, using the
%% bibliography environment described above.  

%%  Contact pnas@nas.edu if you need assistance with your
%%  bibliography.

% Sample bibliography item in PNAS format:
%% \bibitem{in-text reference} comma-separated author names up to 5,
%% for more than 5 authors use first author last name et al. (year published)
%% article title  {\it Journal Name} volume #: start page-end page.
%% ie,
% \bibitem{Neuhaus} Neuhaus J-M, Sitcher L, Meins F, Jr, Boller T (1991) 
% A short C-terminal sequence is necessary and sufficient for the
% targeting of chitinases to the plant vacuole. 
% {\it Proc Natl Acad Sci USA} 88:10362-10366.

%% Enter the largest bibliography number in the facing curly brackets
%% following \begin{thebibliography}

\bibliographystyle{pnas}
%\bibliography{Cage_Form,HS}

%\end{article}
%%%%%%%%%%%%%%%%%%%%%%%%%%%%%%%%%%%%%%%%%%%%%%%%%%%%%%%%%%%%%%%%

%% Adding Figure and Table References
%% Be sure to add figures and tables after \end{article}
%% and before \end{document}

%% For figures, put the caption below the illustration.
%%
%% \begin{figure}
%% \caption{Almost Sharp Front}\label{afoto}
%% \end{figure}

%% For Tables, put caption above table
%%
%% Table caption should start with a capital letter, continue with lower case
%% and not have a period at the end
%% Using @{\vrule height ?? depth ?? width0pt} in the tabular preamble will
%% keep that much space between every line in the table.

%% \begin{table}
%% \caption{Repeat length of longer allele by age of onset class}
%% \begin{tabular}{@{\vrule height 10.5pt depth4pt  width0pt}lrcccc}
%% table text
%% \end{tabular}
%% \end{table}

%% For two column figures and tables, use the following:

%% \begin{figure*}
%% \caption{Almost Sharp Front}\label{afoto}
%% \end{figure*}

%% \begin{table*}
%% \caption{Repeat length of longer allele by age of onset class}
%% \begin{tabular}{ccc}
%% table text
%% \end{tabular}
%% \end{table*}

\end{document}